# Density scaling of the diffusivity in viscous liquids: Identification of the scaling exponent γ with the pressure derivative of the isothermal bulk modulus $(\partial B/\partial P)_T$

**A.N. Papathanassiou (*) and I. Sakellis**

University of Athens, Physics Department, Solid State Physics Section,
Panepistimiopolis, 15784 Zografos, Athens, Greece

*(Dated: 17 Sep. 2009)*

**Abstract**

A density scaled diffusivity function for viscous liquids derived earlier [Phys. Rev. E **79**, 032501 (2009)] is revisited, based on an improved equation of state assuming that the isothermal bulk modulus increases linearly with pressure. Without making any assumption on the interconnection between the scaling exponent and the Grüneisen parameter, we prove that the scaling exponent is identical with the pressure derivative of the isothermal bulk modulus. We further discuss probable interconnection between the scaling exponent and the Grüneisen parameter.

(*) Corresponding author; E-mail address: antpapa@phys.uoa.gr

**Introduction**

I.

A density scaled (reduced) diffusivity equation for viscous liquids was suggested recently [1]:

$$\ln D(\rho^\gamma) \cong -\frac{\Lambda}{\gamma}(\rho^\gamma)^2 - \left\{2\Lambda\left(1-\frac{1}{\gamma}\right)\right\}\rho^\gamma + \left(2-\frac{1}{\gamma}\right)\Lambda - 1 \qquad (1)$$

where D and ρ are the reduced diffusivity and reduced density, respectively, γ denotes the scaling exponent, Λ is a quantity related with the material's fragility. The procedure and assumptions employed to formulate Eq. (1) are presented in detailed [1].

Eq. (1) was based on an *approximate* equation of state (EOS):

$$\rho^\gamma \cong 1 + \frac{\gamma}{B}P \qquad (2)$$

where $B_0$ is the ambient (zero) pressure isothermal bulk modulus. Eq. (2) was derived from the definition of the isothermal bulk modulus

$$B \equiv -(\partial P/\partial \ln V)_T \qquad (3)$$

Recalling that $\rho \equiv m/V$, we get:

$$B = (\partial P/\partial \ln \rho)_T \qquad (4)$$

Further, the scaling constant γ was introduced:

$$B \equiv \gamma(\partial P/\partial \ln \rho^\gamma)_T \qquad (5)$$

Although the introduction of γ is mathematically correct, it is useless and makes difficult the physical interpretation of the formulas, as it will be shown in the following. The solution of Eq. (5) under *either* the assumption of that B is slow varying with pressure *or* under the more realistic assertion that that B(P) increases linearly with pressure

$$B(P) = B_0 + (\partial B/\partial P)_T P \qquad (6)$$

where $(\partial B/\partial P)_T$ is considered constant, to a first approximation, and keeping first order expansion pressure terms, we reproduce the *approximate* EOS determined by Eq. (2).

**Refining the density scaling diffusivity equation**

In this manuscript, we show that the abrupt introduction of γ is useless (although it is mathematically correct) to built up a density scaled diffusivity function and extract a straightforward relation between the scaling exponent γ and $(\partial B/\partial P)_T$. Moreover, we prove that γ is roughly twice the value of the Grüneisen parameter. The above thermodynamic relations mentioned in the Introduction, yield an exact EOS (within the viewpoint of linear variation of B(P)): Eqs. (4) and (6) merge to:

$$\left(\frac{\partial P}{\partial \ln \rho}\right)_T = B_0 + (\partial B/\partial P)_T P \qquad (7)$$

which, after integrating, gives the following equation of state (EOS):

$$\rho^{(\partial B/\partial P)_T} = 1 + \frac{(\partial B/\partial P)_T}{B_0} P \qquad (8)$$

where ρ holds now its reduced value. We note that the only assumption asserted to derive the latter EOS, is that B(P) increases linearly. By comparing Eq. (8) with Eq. (2), it seems that the scaling exponent γ coincide with $(\partial B/\partial P)_T$. The latter stems straightforwardly as soon as the density scaled diffusivity function is constructed.

Let us use Eq. (8), instead of the approximate EOS Eq. (2) employed initially in [1]. Following the pathway used in [1], (in brief: starting from an earlier approximate diffusivity vs pressure equation [3], incorporating the cBΩ elastic solid state point defect model [4] and hosting the feature of fragile viscous liquids that the activation enthalpy is $h^{act} \approx \Lambda kT$, Λ being of the order of 10) and skipping the assumption initially made initially in [1] that the scaling exponent γ is numerically *roughly equal* to the Grüneisen parameter $\gamma_G$ [2] (such assumption is not required in the present proof), we get:

$$\ln D = -\frac{\Lambda \gamma_G}{(\partial B/\partial P)_T^2} \left(\rho^{(\partial B/\partial P)_T}\right)^2$$

$$-\frac{\gamma_G}{(\partial B/\partial P)_T} \left[2\Lambda\left(1 - \frac{1}{(\partial B/\partial P)_T}\right) - 1\right]\rho^{(\partial B/\partial P)_T} \quad (9)$$

$$+\frac{\gamma_G}{(\partial B/\partial P)_T}\left[\Lambda\left(2 - \frac{1}{(\partial B/\partial P)_T}\right) - 1\right]$$

We observe that the scaling exponent of this equation is identical to $(\partial B/\partial P)_T$; i.e.,

$$\gamma = (\partial B/\partial P)_T \quad (10)$$

If we adopt solid state physics to describe the supercooled state, we can recall that, within Stater's approach [5]:

$$\gamma_G = \frac{1}{2}\left(\frac{\partial B}{\partial P}\right)_T - \frac{1}{6} \quad (11)$$

which was later corrected by Dugdale and McDonald [6] and by Shanker *et al* [7]:

$$\gamma_G = \frac{1}{2}\left(\frac{\partial B}{\partial P}\right)_T - \frac{1}{2} \quad (12)$$

From Eqs. (10) and (11) we get:

$$\gamma = 2\gamma_G + \frac{1}{3} \quad (13)$$

while, from Eq. (10) and (12), we get:

$$\gamma = 2\gamma_G + 1 \quad (14)$$

We observe that, depending on how large the value of $\gamma_G$ is, the scaling parameter $\gamma$ may be close to $2\gamma_G$. This is corth to compare with the result $\gamma \approx 2\gamma_G$ derived in [8] by an alternative way.

**Conclusions**

The refined density scaled diffusivity equation (Eq. (9)) is a second order polynomial function with respect to $\rho^{(\partial B/\partial P)_T}$. The *advantages* of the refined scaling function (Eq. (9)) in comparison with its equivalent one (Eq. (1)) [1] are the following:

(i)  it is not essential to incorporate the scaling exponent γ into the EOS
(ii) there is no need to assume that γ is numerically *roughly* equal to $\gamma_G$.

Eq. (9) is qualitatively similar to the scaling function derived in [1] (Eq. (1)) and, indicates that the scaling exponent γ is identical with $(\partial B/\partial P)_T$, which, furthermore, yields probably an interconnection between γ and $\gamma_G$.